\documentstyle[epsf,prl,aps]{revtex}
\begin{document}
\twocolumn[\hsize\textwidth\columnwidth\hsize\csname @twocolumnfalse\endcsname

\title{Extended Moment Formation and Second Neighbor 
Coupling in Li$_2$CuO$_2$}
\author{Ruben Weht
and W.E. Pickett \cite{corresponding}}
\address{Department of Physics,
University of California, Davis CA 95616}
\date{\today}
\maketitle
\begin{abstract}
Comprised of ferromagnetic edge-sharing CuO$_2$
chains that order in antialigned fashion at T$_N$=9 K, Li$_2$CuO$_2$ is found
from local spin density calculations to display several surprising
characteristics: (1) the ordered moment/f.u. of 0.92 $\mu_B$ is the largest
for any low dimensional cuprate system, in agreement with experiment,
(2) 40\% of this moment lies on the neighboring O ions,
making it the largest oxygen moment yet reported, and (3) the
second neighbors couplings are larger than nearest neighbors couplings. 
All of these phenomena arise naturally due to a
well defined effective $d_{yz}$ type orbital that includes very strong
O $p_{\sigma}$ character.  We interpret the large moment as surviving
reduction by quantum fluctuations due to extension caused by the
$d-p$ hybridization.

\end{abstract}
]

Low dimensional (1D or 2D) spin S=$\frac{1}{2}$ materials are 
critical systems displaying a
variety of phenomena in crystals. These range from spin gap behavior in
CaV$_4$O$_9$\cite{taniguchi} to anomalous spin
physics in the one-dimensional spin chains and ladders.\cite{dagrice} 
Many of these characteristics
depend crucially on specific structural or chemical bonding 
features.\cite{wepcavo}
The magnetic coupling is particularly sensitive, with simple nearest
neighbor (nn) exchange varying from large and 
antiferromagnetic (AF) to small and ferromagnetic (FM) 
when the metal-oxygen-metal angle
$\phi$ varies from 180$^{\circ}$ to 90$^{\circ}$.  
Weak intrachain couplings
and competing exchange couplings can lead to frustration, magnetic
ordering,  spin gap behavior or spin-Peierls phase formation.  

Copper oxides play a very
important role due to the various possibilities of linking their
fundamental unit, a (often slightly distorted) CuO$_4$ square.
Three general types of arrangement can be
found in these systems, classified in terms of the oxygen squares 
sharing corners (as in the high T$_c$ planar compounds
and Sr$_2$CuO$_3$), 
edges (as in CuGeO$_3$, La$_6$Ca$_8$Cu$_{24}$O$_{41}$, and Li$_2$CuO$_2$),
\cite{edgeshare}, or both (as in SrCuO$_2$).  In
edge sharing CuGeO$_3$, for example, there is 
moderately strong antiferromagnetic (AF)
nn coupling (J$\approx$150 K),\cite{cugeo3} 
but a spin-Peierls transition occurs only at 14 K.
The typical example of corner-sharing chains is Sr$_2$CuO$_3$, 
where the system
orders antiferromagnetically\cite{sr2cuo3}
with a low transition temperature of 5 K and a
small induced magnetic moment (0.06 $\mu_B$) in spite of
very large interactions between ions
(J$\approx$2200 K). Despite considerable progress a clear understanding
of the magnetic behavior of the Cu$^{2+}$ ion in several
regimes is still lacking.

Here we report a new aspect of spin behavior in edge-sharing systems
revealed by spin-polarized local density approximation (LDA) studies
of a 1D S=$\frac{1}{2}$ system Li$_2$CuO$_2$.  In this compound
neutron scattering\cite{structure} indicates 
three dimensional AF ordering at 9 K arising from the antialignment of FM
chains.  The experimental
moment of 0.9 $\mu_B$ per cell was attributed
completely to the Cu ions.
Based upon experience in the undoped two-dimensional cuprates
({\it viz.} La$_2$CuO$_4$) where LDA is unable to obtain any moment
whatsoever on the Cu ion\cite{rmp}, 
it might seem that LDA is unlikely to produce a magnetic Cu ion or
an insulating system. Due to this problem with
corner-sharing Cu-O planes, very few spin-polarized
calculations on cuprate compounds have been reported.  
However, we find
that LDA predicts Cu in Li$_2$CuO$_2$ to be robustly magnetic, allowing
us to obtain the relative energies
and electronic properties of the system with FM chain with
both AF and FM coupling between chains, the AF chain, and the unpolarized
(PM) system taken as reference.  The oxygen ions play
a fundamental role in the band dispersion and the magnetism, and carry a
magnetic moment approaching 0.2 $\mu_B$ per atom, 
the largest O moment yet reported.  Previous reports of O moments lie in
the 0.02-0.10 $\mu_B$ range.\cite{LSMO}

\begin{figure}
\epsfxsize=8.9cm\centerline{\epsffile{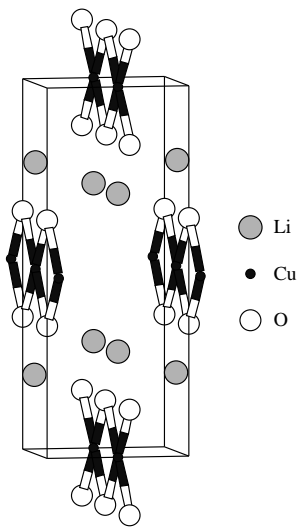}}
\caption{Crystal structure of Li$_2$CuO$_2$, for the
AF unit cell containing two CuO$_2$ ribbons.
\label{Fig1}}
\end{figure}

Li$_2$CuO$_2$ is orthorhombic and belongs to the category of
edge-sharing compounds, with one-dimensional CuO$_2$ ribbons carrying
the Cu chains along the $b$ axis,
arrayed in a body-centered fashion in the $a-c$ plane (Fig.~\ref{Fig1}). 
The Cu-O-Cu angle $\phi$ = 94$^{\circ}$, 
intermediate between the two other edge-share compounds with very
different characteristics: GeCuO$_3$ (spin-Peierls, $\phi = 99^{\circ}$) and
La$_6$Ca$_8$Sr$_{24}$O$_{41}$ (FM, $\phi = 91^{\circ}$).
Distances between two Cu ions, 
2.86 \AA~along the chain, 3.65 \AA~in the $a-b$ plane 
and 5.23 \AA~in the diagonal direction, do not reflect
the relative coupling strengths, as we explain below.

Calculations were done using the linearized augmented plane
wave (LAPW) method,\cite{wien} which makes no shape approximations for the
density or potential. The sphere radii
used in fixing the LAPW basis were chosen to be 2.00 a.u for Cu and Li
and 1.65 for O. Local orbitals (Cu $3p$; Li $1s$; O $2s$)
were added to the basis set for extra
flexibility and to allow semicore states to be treated within the same
energy window as the band states. The plane wave
cutoff corresponded to energy of 23.5 Ry resulting in 640
LAPWs per formula unit. Self-consistency was carried out on k-points meshes
of 512 points in the Brillouin zone for the compounds which need a single
unit cell calculation and 256 points when we considered an 

\noindent
AF arrangement for the chains.

The paramagnetic system has an odd number of electrons per unit cell
and thus is
metallic.  There is a single band in the range of 1 eV around the Fermi
level split off from the rest of the $p-d$ band complex, similar to what
was found in the other CuO$_2$ edge-sharing compounds CuGeO$_3$
\cite{matth} and NaCuO$_2$.\cite{djs}. This isolated band shows up as
two bands in Fig.~\ref{Fig2}, where a doubled cell with two chains has been
used for comparison with the AF bands (see below).  The analysis 
of the partial density of states shows that only Cu
d$_{yz}$-O $p_{\sigma}$ (the $p_y \pm p_z$ combination directed towards
the Cu site) are present in this band.  

The geometry of this CuO$_2$ edge-sharing chain leads to a
simple description of the important band, which we expect (and find) 
to involve an antibonding combination of Cu $d$ and O $p$
orbitals.  The atomic orbital basis in a primitive cell can be chosen 
as the Cu $d$ orbitals, the $\sigma$-type
O $p_{\sigma}$ orbitals on each of the {\it four} neighboring O atoms
which strongly overlap the $d_{yz}$ orbital, and half of the
out-of-plane $p_x$ orbitals that are non-bonding 
and lie well below E$_F$ (as do all $d$ orbitals except $d_{yz}$).  
The $p_{\pi}$ orbitals in the $y-z$ plane of 
the ribbon are
$p_{\sigma}$ with respect to a neighboring Cu and
belong to the next unit cell.  The $d_{yz}$ and the four $p_{\sigma}$
orbitals can be decomposed into five hybridized combinations:
one bonding combination
${\cal D}_{yz}$ and one antibonding combination ${\cal D}_{yz}^*$
of $d_{yz}$ and the combination of the four $p_{\sigma}$ 
orbitals with d$_{yz}$ symmetry, and three other 
$p_{\sigma}$-only combinations of lower symmetry. ${\cal D}_{yz}$ and
${\cal D}_{yz}^*$ are split strongly, leaving ${\cal D}_{yz}^*$ at the 
Fermi level and ${\cal D}_{yz}$ 5 eV below, as shown in the 
density of states plot in Fig.~\ref{Fig3}.  The others lie 
around -4 eV below E$_F$ and are not of interest.  
The general behavior of the coupled
$d_{yz}-p_{\sigma}$ cluster can be modelled with $\varepsilon_d=-1.5,
\varepsilon_p=-4, (dp{\sigma})=\pm 1.15, (pp{\sigma})=
0.25,$ (all in 

\begin{figure}
\epsfxsize=8.9cm\centerline{\epsffile{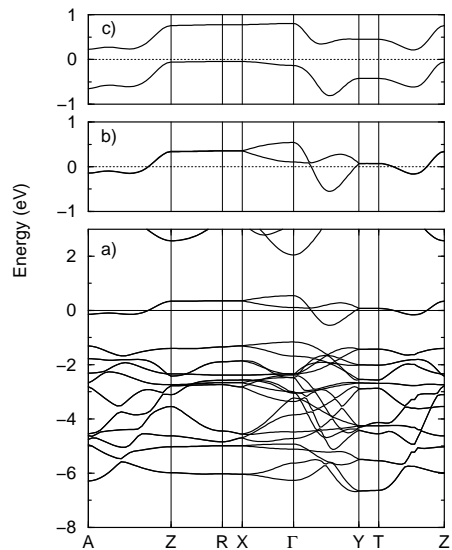}}
\caption{Band structure of a) paramagnetic Li$_2$CuO$_2$, 
b) a zoom of the region
around the Fermi energy, and c) for antiferromagnetic 
alignment of FM chains. $\Gamma$, X, Y and Z points correspond to the
standard denomination,  ${\it A = (\pi,\pi,\pi)}$, ${\it R=(\pi,0,\pi)}$
and ${\it T= (0,\pi,\pi)}$; all in the double cell Brillouin zone.
\label{Fig2}}
\end{figure}

\noindent
eV).  For these parameters the ${\cal D}_{xy}^*$ density
is 70\% on the Cu and 30\% on the four O ions.

The active orbital ${\cal D}_{yz}^*$ shown (schematically)
as $| {\cal D}_{yz}^* |^2$ on next nearest neighbors in 
Fig.\ref{Fig4},
is an effective $d_{yz}$-type 
orbital centered on each Cu ion but extending strongly to the
neighboring O sites.  Symmetry allows direct ${\cal D}{\cal D}\pi$
overlap, and therefore hopping amplitude $t_{\pi}$ along the
chain.  Due to its parentage, however, it is clear that the
main contribution to the overlap arises from the O ion region.
If the O quadrilateral were perfectly square ($\phi=90^{\circ}$)
the O $p_{\sigma}$
orbitals directed toward the two neighboring Cu ions would be precisely
$p_y \pm p_z$.  These combinations are orthogonal, so $t_{\pi}$
reduces to direct d-d overlap and 
will be very small.  When the Cu-O-Cu angle is
not exactly 90$^{\circ}$, the $p_{\sigma}$ orbitals are
no longer orthogonal and the overlap (and $t_{\pi}$) 
increases.

The effective Hamiltonian therefore reduces to a single orbital
(${\cal D}_{yz}^*$) per cell.  The dispersion of the ${\cal D}_{yz}^*$
band cannot be fit simply by nn hopping $t_{\pi}=t_1$ along the chain
and $t^{\prime}_1$ between neighboring chains (in roughly the 
$\hat x \pm \hat z$ directions), but requires as well both next nearest 
neighbor (nnn) hopping terms $t_2$ and $t^{\prime}_2$.  Consideration of the
${\cal D}_{yz}^*$ orbitals on second neighbors indicates why this is
so: Cu-O-O-Cu coupling along the ribbon becomes important because
of O-O coupling and because
the nn hopping is so small, and similarly for interchain hopping
along the diagonal.  The manner of second neighbor overlap
along the chain is clear in Fig.\ref{Fig4}.
The values 
\begin{eqnarray*}
nn: ~~~~ t_1&=&-63 ~meV ~ , ~~ t_1^{\prime}=-16 ~meV,  \\  
nnn:~~   t_2&=&-94 ~meV ~ , ~~ t_2^{\prime}=~~44 ~meV,          
\end{eqnarray*}

\begin{figure}
\epsfxsize=9.0cm\centerline{\epsffile{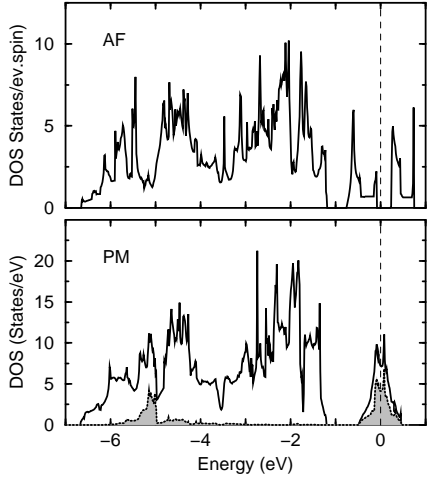}}
\caption{Density of states for the paramagnetic system 
and for AF alignment of the chains.
In the PM case the partial density of states for 
the Cu d$_{yz}$ is shown as the shadowed region, illustrating
the 5 eV splitting of the ${\cal D}_{yz}$ and ${\cal D}^*_{yz}$ bands.
\label{Fig3}}
\end{figure}

\noindent
(note that second neighbors values
are larger than nearest neighbors values), provide an excellent
fit to the dispersion except near the ($\pi,\pi,k_z$) line (not shown) where
some additional 
interaction would be required to obtain a perfect fit.

This emergence of an effective
single-band system with a simple, strongly hybridized orbital
is one of the cleanest in
Cu-O systems.  It implies that the (ferro- or antiferro-)
magnetism of this compound should be interpreted in terms of 
this extended ${\cal D}_{yz}^*$ orbital instead of the Cu $d_{yz}$
orbital; likewise, correlation effects must involve an on-site
repulsion U$_{\cal D}$ rather than U$_d$ and therefore should be
smaller than might have been anticipated.  Likewise, the
exchange splitting of this band will be significantly
different from that of the other $d$ orbitals.

The spin wave dispersion curves have been measured by Boehm 
{\it et al.}\cite{Jli2cuo2}  Their fit requires important nnn
exchange couplings, consistent with our finding that nnn hopping
is essential to account for the dispersion.  Although our hopping
parameters are small, the exchange constants they obtain are very
small, $\vert$J$\vert \le$0.4 meV.  Application of the superexchange
expression J=4$t^2$/U leads to unphysically large estimates of
U$\approx$70-100 eV, indicating that
exchange coupling requires more careful consideration. 

Relative to the paramagnetic system, the FM state gains 70 meV/f.u.
AF chains lead to a slight lowering by another 1.5 meV/f.u., but 
the phase with oppositely
aligned FM chains is lower still (consistent with observation),
80 meV/f.u. below the unpolarized state.  
The density of states for these phases are
shown in Fig.\ref{Fig2}, revealing an
insulating gap of 0.32 eV.  
As usual, correlation effects will widen the gap, 
but we are not aware of any 

\begin{figure}
\epsfxsize=9.0cm\centerline{\epsffile{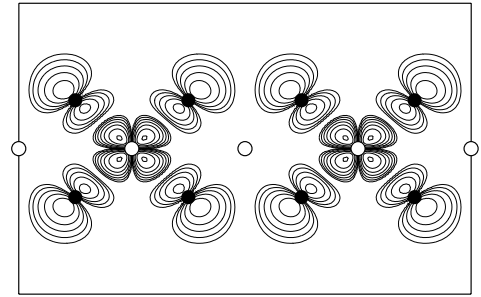}}
\caption{
Illustration of the overlap of two densities 
$| {\cal D}_{yz}^* |^2$
centered on second neighbors Cu ions. Cu and O sites are denoted
by empty and filled circles respectively.
Model Gaussian $d_{yz}$ and $p_{\sigma}$ orbitals were used for this plot.
\label{Fig4}}
\end{figure}

\noindent
published data.

The moment of 0.92 $\mu_B$ per CuO$_2$
unit is roughly 60\% on Cu and 40\% on O ions.  This transfer
of magnetic moment from a transition metal ion to a ligand ion (almost
0.2 $\mu_B$ on each O)
is to our knowledge the strongest yet reported in a transition
metal oxide compound\cite{k2ircl6}.  

\begin{figure}
\epsfxsize=9.0cm\centerline{\epsffile{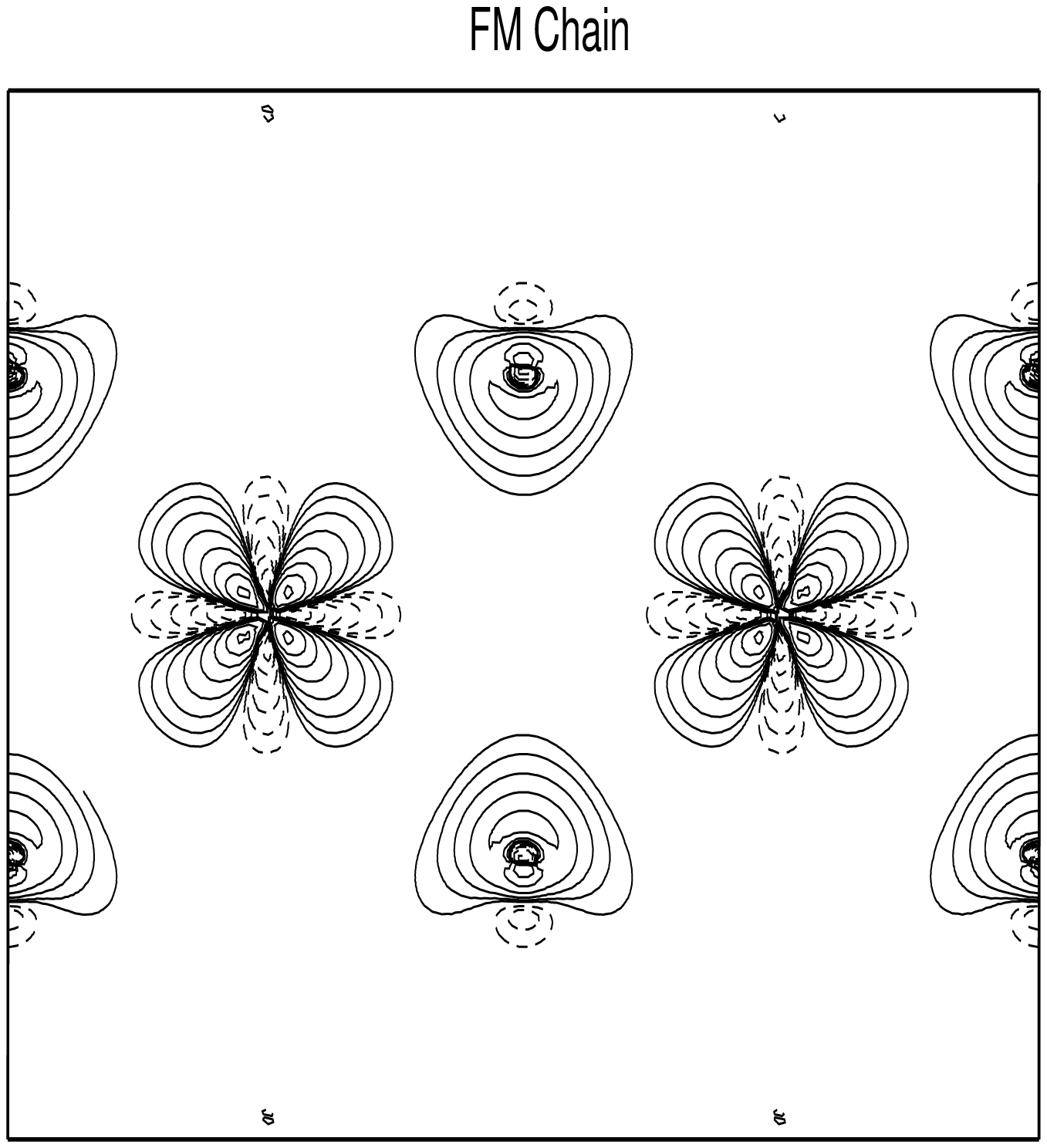}}
\epsfxsize=9.0cm\centerline{\epsffile{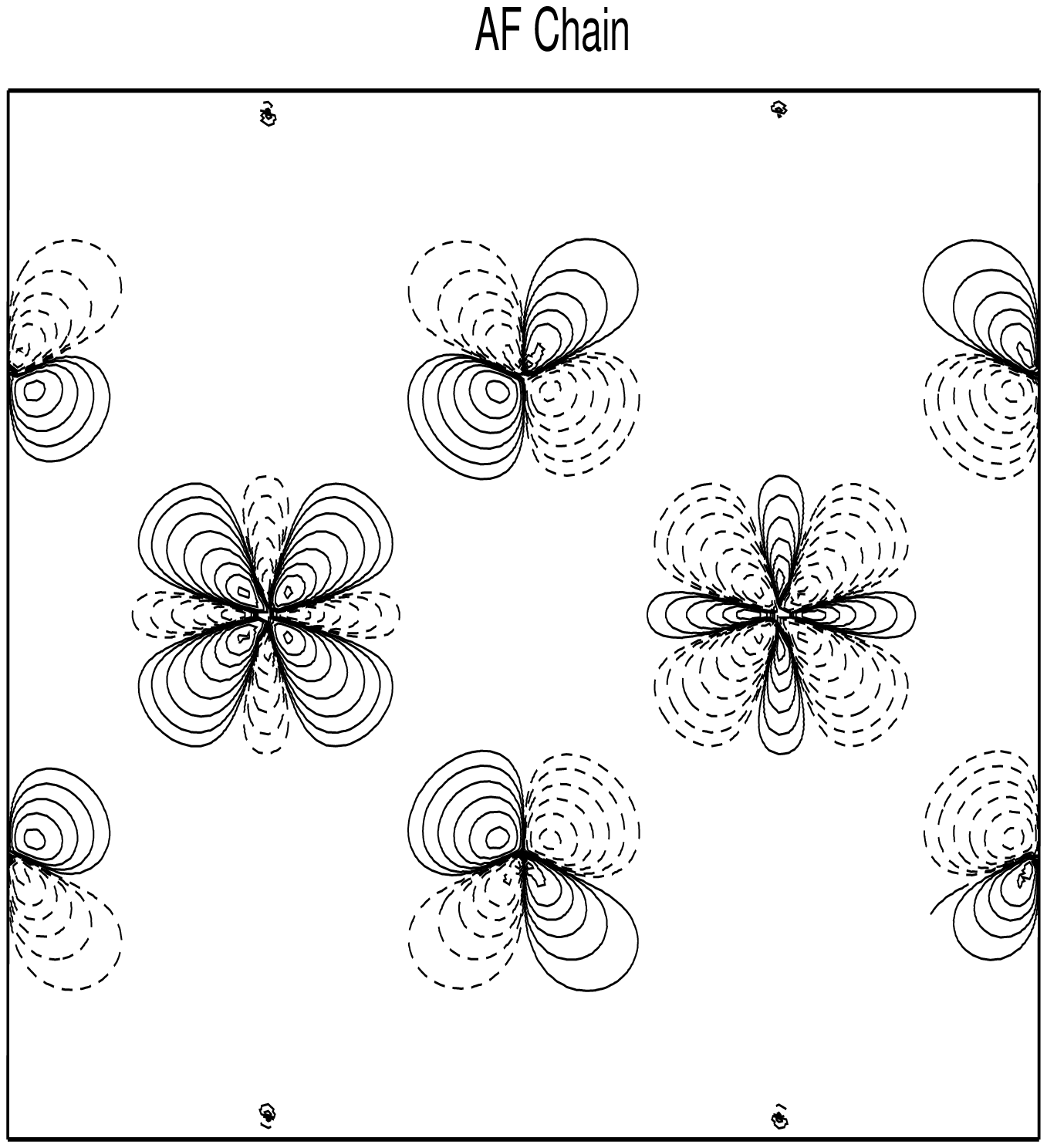}}
\caption{Exchange potential for ferromagnetic and antiferromagnetic 
Cu-O chains.
Note for the AF case the close resemblance to the 
${\cal D}_{yz}^*$ orbital density. The solid and dashed lines
represent negative and positive values respectively.
\label{Fig5}}
\end{figure}

Previous reports of moments lie in
the 0.02-0.10 $\mu_B$ range.\cite{LSMO} 
To illustrate the importance of the O sites, we display the exchange potential
$V_{\uparrow}-V_{\downarrow}$ for the ferromagnetic and antiferromagnetic 
chains in Fig.~\ref{Fig5}.
There is a clear similarity, especially for the AF chain,
to the ${\cal D}_{yz}^*$
density in Fig.\ref{Fig4}.
Two other features should be noted: 
the exchange potential on O is comparable to that
on Cu, and the exchange potential is of predominantly one sign for 
the entire ${\cal D}_{yz}^*$
orbital for both FM and AF ordered chains.
Unlike the strong dependence of O moment on the magnetic order,
the size of the moment on the Cu ion itself (0.50-0.55 $\mu_B$)
is essentially independent of the overall magnetic ordering.

Antialignment of FM chains leads to the insulating band structure shown in
Fig. 2(c).  As expected, the two bands are described well
by the eigenvalues of the system
\begin{eqnarray*}
\left(
\begin{array}{cc}
 t_{1,1}(k) + \frac{1}{2}\Delta &  t_{1,2}(k) \\
 t_{2,1}(k)    & t_{2,2}(k) - \frac{1}{2}\Delta \\
\end{array}
\right),
\end{eqnarray*}
where $t_{1,1}=t_{2,2}$ contains all intrachain hopping, $t_{1,2}=
t_{2,1}$ contains the interchain hopping given above, 
and $\Delta$=0.8 eV is the exchange splitting evident in Fig.2(c).  
Since $\Delta$ is much greater than all of the
hopping amplitudes and in fact similar to the PM bandwidth, the 
exchange coupling between the antialigned chains results in very
narrow band splits by $\Delta$.  

Although the chains are nominally fully polarized,
the coupling reduces the net moment to $0.92 \mu_B$/f.u.
This total moment is in excellent agreement with neutron scattering
results,\cite{structure} although the moment there was 

\noindent
attributed solely
to Cu.  This is a remarkably large value for a quasi-1D S=$\frac{1}{2}$
chain, where quantum fluctuations should be large.  The strong nnn
coupling that we have uncovered account for the observation 
qualitatively: coupling between the ${\cal D}^*_{yz}$ effective
orbitals, and therefore the 
spins, is really three dimensional, hence
quantum fluctuation effect are vastly reduced.  

Recently two similar compounds with one-dimensional chains have been
reported. 
Both Sr$_{0.73}$CuO$_2$ and Ca$_{0.85}$CuO2$_2$ have CuO$_2$ ribbons and order
magnetically.\cite{srcuo2} 
Comparison with Li$_2$CuO$_2$ is difficult, however, because the strongly 
differing doping level and the structural disorder can lead to large changes in the magnetic behavior.
Recent work on the Ca$_{2+x}$Y$_{2-x}$Cu$_5$O$_{10}$\cite{hayashi}
is suggestive, 
since susceptibility measurements are interpreted as indicating that each
extra doped hole creates a (non-magnetic) Zhang-Rice singlet.
Our results for Li$_2$CuO$_2$ suggest that for the edge-sharing chains each
extra hole occupies a ${\cal D}_{yz}^*$ orbital and
therefore is weakly coupled to neighboring Cu ions. 
In this case the Cu ion and four neighboring p$_\sigma$ orbitals would be
non-magnetic, rather than having a Cu spin be compensated by
neighboring O spins. Further experiments will be necessary
to test our picture.

To summarize, we have found that LDA provides a consistent picture
of the magnetic properties and insulating character of the 
quasi-one-dimensional
antiferromagnet Li$_2$CuO$_2$.
Due to the formation of a strongly hybridized, and energetically
isolated, combination of $d_{yz}$ and $p_{\sigma}$ orbitals, a
large moment is transferred to the O ions.  A simple single-band
system results, but one in which second neighbors
coupling exceeds nearest neighbors coupling and the electronic and magnetic
behavior is three dimensional.

We acknowledge stimulating conversations with G. Shirane
and a preprint of Ref.\onlinecite{Jli2cuo2} from A.Zheludev.
This research was supported by Office of Naval Research
Grant No. N00014-97-1-0956.

\end{document}